# Software as a Service: Analyzing Security Issues

Pushpinder Kaur Chouhan, Feng Yao, Suleiman Y. Yerima and Sakir Sezer
Centre for Secure Information Technologies
Queen's University of Belfast, Northern Ireland, UK

*Abstract*—*Software-as-a-service (SaaS) is a type of software service delivery model which encompasses a broad range of business opportunities and challenges. Users and service providers are reluctant to integrate their business into SaaS due to its security concerns while at the same time they are attracted by its benefits. This article highlights SaaS utility and applicability in different environments like cloud computing, mobile cloud computing, software defined networking and Internet of things. It then embarks on the analysis of SaaS security challenges spanning across data security, application security and SaaS deployment security. A detailed review of the existing mainstream solutions to tackle the respective security issues mapping into different SaaS security challenges is presented. Finally, possible solutions or techniques which can be applied in tandem are presented for a secure SaaS platform.*

## I. INTRODUCTION

Software business delivery models have changed dramatically during the last decade. It has evolved gradually from the traditional on premise model to the current off-premise model which is also termed as Software-as-a-service (SaaS). SaaS model delivers web-based applications over the Internet. The software is hosted at the providers' site and all the maintenance operations become the providers' responsibility.

Through exploiting SaaS, users can pay a per usage subscription fee without investing a huge amount of money to install and maintain necessary software and hardware. Users can access the service anywhere and anytime in the world as long as they have a device with internet access. Users are guaranteed with the latest version of software usage without bothering with updates. For SaaS service providers, they are motivated by an on-going revenue to focus on improvement of their software. The multi-tenancy and virtualization technologies inherent in SaaS brings the service providers maximized resource utilization and centralized management.

A typical SaaS application is offered either directly by the provider or by an intermediary party called an aggregator, which bundles SaaS offerings from different providers and offers them as part of a unified application platform. A general SaaS mechanism is presented in Fig. 1. The emergence of SaaS as an effective software-delivery mechanism creates an opportunity for IT departments to change their focus from deploying and supporting applications to managing the services that those applications provide.

Despite the advantages of SaaS, security issues and challenges still exist for developing a fully-fledged secure implementation of SaaS. Once the weaknesses are identified, appropriate countermeasures regarding secure data protection, secure web application design and secure virtual environment should be implemented jointly in the right way to maintain a high-level, multi-layer security framework to ensure privacy and data protection for users. The in-depth analysis and classification of existing solutions in this article can be leveraged by the wider research community and industry in developing their own SaaS strategies.

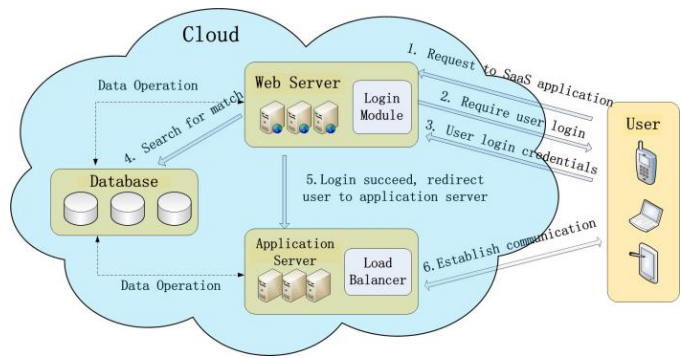

Fig. 1: SaaS Mechanism Overview.

The organization of article is described as follows. In section II, we discuss SaaS integration into different environments. Section III provides a brief overview of SaaS security challenges. Section IV analyzes and summarizes previous work aimed at addressing various SaaS security issues. Section V presents possible security solutions for SaaS and explains how these solutions could enhance security against the vulnerabilities of the current constituent technologies'. Finally, the conclusion is given in section VI.

## II. SOFTWARE-AS-A-SERVICE IN DIFFERENT ENVIRONMENTS

SaaS is a fundamental component of the cloud computing architecture. Similarly, mobile cloud computing incorporates SaaS indirectly through cloud computing. SaaS software de-livery business model can be incorporated in Software Defined Networking (SDN) and Internet of Things (IOT) in a similar manner as cloud computing is being utilized by mobile service providers through Mobile Cloud Computing (MCC).

### A. Software-as-a-Service in Cloud Computing

Cloud computing enables ubiquitous, convenient, on-demand network access to a shared pool of configurable computing resources that can be rapidly made available and released with minimal management effort according to service users requirement [1]. Implementation of SaaS logic in cloud computing is shown in Fig. 2

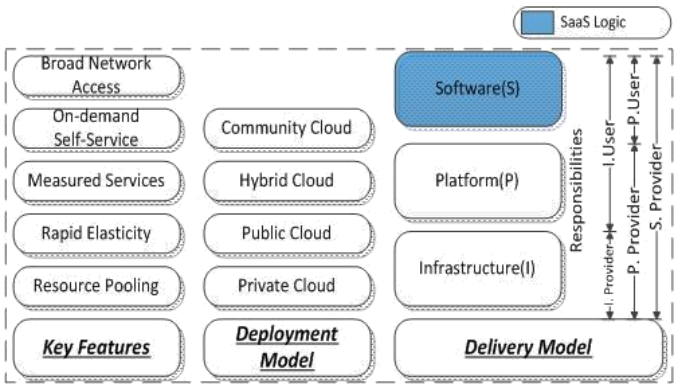
Fig. 2: Conceptual diagram to define Cloud Computing.

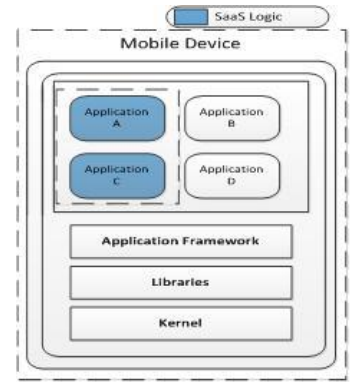
Fig. 3: SaaS in Mobile Cloud Computing.

Cloud computing hosts services defined as "X-as-a-service", where X includes infrastructure (I), platform (P) and software (S). IaaS is the foundation of all cloud services, with PaaS build upon IaaS, and SaaS in turn build upon PaaS. Cloud computing delivery model separates the responsibilities among the cloud provider and cloud user, depending on which depth the cloud is being demanded by the cloud users. The lower down the stack the cloud provider is, the more security responsibilities users need to take on for implementing and managing the hardware, running environment or applications. SaaS providers take extra charge of developing and designing applications securely and properly along with the management of metadata, data and associated APIs of application for end-users.

B. Software-as-a-Service in Mobile Cloud Computing

SaaS enables new types of mobile services and facilitates mobile users to take full advantage of cloud computing. SaaS in the mobile environment can be viewed as a component of mobile cloud computing which enables mobile access to applications and information via the internet and at the same time benefit from rapid provisioning of on-demand elastic services.

Mobile cloud computing (MCC) refers to an infrastructure where both the data storage and data processing happen outside of the mobile device into the cloud. These centralized applications are then accessed over wireless connections based on a thin native client or web browser on the mobile devices.

SaaS is essentially the main driver of MCC (Fig. 3). Clearly, mobile computing derives all the benefits discussed in section I when extended with the SaaS model. In addition, SaaS has unique benefits to mobile environments which will rapidly increase the emergence of mobile based SaaS applications of a wide variety to cater for social, business, education, leisure, information and entertainment needs of mobile users. SaaS provides the opportunity to overcome the limitations of battery life, processing power, storage and memory requirements to provide accessible and secure services to cater for these areas that users have come to expect to be able to access in real-time and while on the move via their smartphones and other mobile devices.

C. Software-as-a-Service in Software Defined Networking

Software Defined Networking (SDN) decouples the control and data planes, logically centralizes the network intelligence/state and abstract the underlying network infrastructure from the applications.

The basis of SDN is separation of layers, which in its most simplistic form allows software to run separately from the underlying hardware. Note that one of the four key features of SDN [2] is "Programmability" of the network by external applications. SaaS being an approach to deliver software, this key feature of SDN can be enhanced by the use of SaaS (shown in Fig. 4).

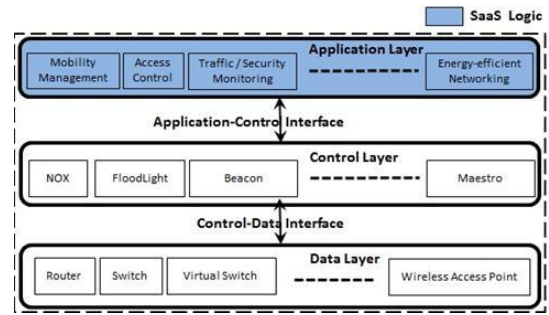
Fig. 4: SaaS in Software Defined Networking.

SaaS can introduce flexible and modular software-based intelligence that can play a major role in changing the way networks are built and operated as part of SDN thus accelerating new service creation to permit network operators to customize and innovate. The separation of the traffic control function from the network hardware in SDN can be expanded through SaaS. SDN with SaaS implementation provide a scope of delivering applications for future network on a pay-as-you-go basis. This can help service providers to sell Network-As-A-Service thus benefiting the user with security, scalability, and manageability.

To incorporate the benefits of SaaS and virtualization technologies in networking field, Network Function Virtualization (NFV) has been formalized by ETSI in 2012. NFV virtualize network functions previously carried out by the proprietary dedicated hardware. NFV decouples network functions such as NAT, firewalling, Intrusion detection, DNS and caching etc. NFV builds virtualized network functions whereas SDN builds a controller that is external to the network, builds a logical network as an overlay over the Internet and adds functionality implemented as conventional software. Network functional applications such as energy-efficient networking, security monitoring and access control for operation and management of the network and services provided by the network operators can be deployed using SaaS.

D. Software-as-a-Service in Internet of Things

The term Internet of Things (IOT) was first introduced by Kevin Ashton in 1999 [3]. However, it has become more popular in recent years. IOT in this article is considered with the 'Coordination and support action for global RFID-related activities and standardization' (CASAGRAS) [4] definition: "A global network infrastructure, linking physical and virtual objects through the exploitation of data capture and communication capabilities. This infrastructure includes existing and evolving Internet and network developments. It will offer specific object-identification, sensor and connection capability as the basis for the development of independent cooperative services and applications. These will be characterized by a high degree of autonomous data capture, event transfer, network connectivity and interoperability".

The development of IOT can mainly be attributed to technologies such as: RFID technology, Sensor technology, embedded technology and the progress of network technologies. IOT dedicate on connecting everything we use into the network. SaaS, being contemporary widely adopted software application delivery service, can possibly play an important role in IOT.

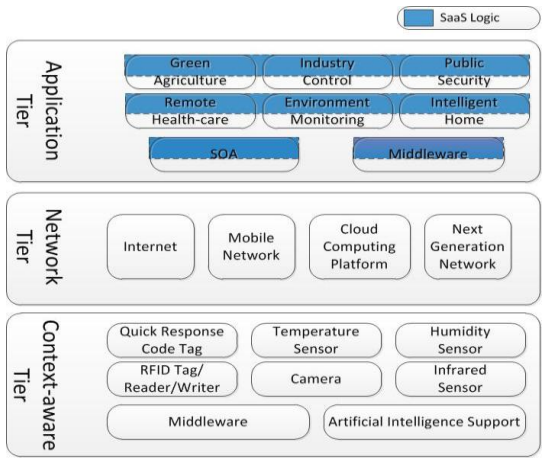

Fig. 5: SaaS in Internet of Things.

As shown in Fig. 5, in order to fulfil various goals such as intelligent home, remote health-care and public security, various agent software should be selected and jointly employed. Some of these could be residing on a SaaS platform. Deploying multifarious agent software on the SaaS platform grants users an easy access to software on a pay-as-you go basis thus enabling a more cost effective IOT deployment model.

III. SOFTWARE-AS-A-SERVICE SECURITY CHALLENGES

The enormous growth of off-premise application services has changed the way the application services are delivered and brings significant benefits and convenience to the software providers and users. However, as more and more individuals and enterprises deploy their applications in the SaaS model, concerns about the security and privacy of their information and reliability of the service is becoming the focus of attention. In this section, SaaS security challenges [5] are presented in 3 main groups which are data security, application security and deployment security.

A. Data Security

Data is one of the most important assets for users which must be kept secure. In SaaS scenario, data resides in the database which is outside the boundary of the enterprise and depends on the provider for proper security measures. Since multi-tenancy through virtualization is a major feature for SaaS, SaaS providers are questioned if they can provide isolated environment for each user in which none of them can see each other's data without permission. SaaS consumers have no idea "how strong the access control system is?" to prevent unauthorized access. The transmission channel between SaaS providers and users is not considered always secure. Some providers only use SSL during login session, leaving user data unprotected in following sessions. In addition, data backup and recovery should be taken into consideration by SaaS provider to minimize the impact of accidents.

B. Application Security

SaaS applications are mostly used and managed over the web. They are presented to users in a browser. This makes it inevitable to confront the security challenges such as SQL injection, Cross-site scripting and Cross-site Request Forgery. API is the backbone of SaaS platform which aims to deal with heterogeneity and allow automation of common process that interact with services running on another machine. The benefit of API to SaaS is significant, but it is also plagued with security issues. Poorly coded APIs can be easily abused or misused by an attacker. Web service is a software system designed to support machine-to-machine interaction over network. WSDl and SOAP are used largely by SaaS applications, typically conveyed using HTTP with an XML serialization. These techniques, however, are found to be vulnerable to various attacks [6], [7] such as XML wrapping attacks and WSDL scanning. Denial of Service (DoS) and Distributed Denial of Service (DDoS) attack is another security concern which can bring down the SaaS services for hours if there is no efficient countermeasures in place. In addition, rampant malware concealed in the SaaS platform is another threat targeting the user. As people tend to access SaaS applications via mobile devices, substantial amount of malware began to surface targeted at mobile devices, especially on Android-based devices.

C. Software-as-a-Service Deployment Security

Virtualization refers to the act of creating different instances on hardware and on each instance a guest OS is installed. Nowadays, SaaS is largely built on virtualization technology to provide multi-tenancy. However, the vulnerability and weakness of virtualization [8] affects the SaaS security. Even rootkit attackers can gain access to the running instance hosted on the hypervisor, thus, can monitor another VM's resources and CPU utilization, read inbound and outbound traffic of another instance and shut down any instances. Virtual network in SaaS can also degrade the security level [9].

## IV. SECURITY ANALYSIS IN SOFTWARE-AS-A-SERVICE

The basic properties of information security are CIAAN (Confidentiality, Integrity, Availability, Authenticity and Non-repudiation). As SaaS deals with application, data and its information, similar level of security has to be provided to SaaS. Thus, general solutions available for data security, application security and virtualization security can be applicable with little or no modification to provide solution to SaaS security challenges. This section will provide some of the existing SaaS solutions and techniques which can be used in securing SaaS deployment.

Lots of data security analysis has been done in [10]–[13]. However, proposed solutions targeting at up-to-date data security challenges remain insufficient. Moreover, existing solutions are limited by some factors which need to be taken into consideration before being applied to SaaS. The concept of data security in SaaS environment traverses several areas, which includes data confidentiality, data availability, data integrity, data authorization, data backup and recovery, data transfer etc. These data security issues are also addressed in [14], [15] A framework is proposed in [16] through which data confidentiality, availability and integrity is achieved by utilizing SSL encryption, MAC (Message Authentication Code), searchable encryption and division of data. But it does not deal with the issues such as false certificate authorities and fabricated certificates. Moreover, the encryption of all the data incurs a considerable overhead. [17] uses homomorphic token with distributed verification of erasure-coded data to ensure data storage security and locate the server being attacked. The homomorphic token technique is highly efficient and resilient to Byzantine failure, malicious data modification attack and even server colluding attacks. However, in this approach data error location needs to be addressed further. To address data access control issues, [18] presented a scheme that allows querying encrypted database in a multi-user setting, the owner of database can grant different data access privilege to different users. However, authors have proved their approaches efficiency on SQL-like database structure only. To achieve high-level data access control scheme, [19] proposed a generic security management framework that enables cloud storage providers to define and enforce flexible security policies. This approach firstly defines templates of various attacks and map them into security policies, then based on these policies a security violation detection engine is developed to search for recorded user actions that match the template events defined by the policy.

Asigra [20] introduces a new scheme for backup solution called agentless backup and recovery. Unlike agent backup solution which requires a backup agent installed on the machine that needs to backup, agentless backup usually installs client software on one server which can remotely log into a target backup system and transmit the data to the central storage location. The advantage of agentless backup is convenience. So, service providers do not have to manage agents installed on each client machines. The disadvantages of this approach are that agentless backup is still not mature enough and the server with agentless client software might open another attack vector for the hackers. The authors in [21] proposed a high-availability and integrity layer (HAIL) for cloud data storage. It is a remote data integrity checking protocol to ensure data integrity and availability through an interleaving of different types of error-correcting layers which leverage dispersal code, server code and aggregation code. The limitation of this approach is that it can only provide assurance for static data. Trusted Third Party [22] ensure the authentication, integrity and confidentiality of involved data and communications in cloud environment by utilizing public key infrastructure operating in concert with LDAP and SSO. The limitation of Trusted Third Part is that it does not address issues such as data availability, data backup and recovery.

Application security is still a hot and important topic which has been analyzed in [6], [10], [12]. Despite the large quantity of research work that has been done on this issue, fine-grained solutions put forward to tackle application security issues are relatively less. The issues and requirement associated with the development of web-based SaaS application are presented in [23], [24]. It encapsulates issues regarding state-less architecture, published APIs and multi-tenancy feature of SaaS. It is critical that SaaS applications incorporate strict authentication mechanisms to authenticate their users before granting access. A user authentication framework is proposed in [25] where user identity is verified before user gets access to the resource required. It provides identity management, mutual authentication, session key establishment between user and service provider. The advantage of this framework is that it can repel many famous attacks such as replay attack, MIMT attack and DoS/DDoS attack. But some formal security proofing techniques need to be provided to uphold this approach and this approach is built on some preconditions. Cognitive approach is proposed in [26] i.e. the devices should act more like humans when making a security decision. Devices should possess identifying quality by utilizing the existing machine learning approach such as RBF (Radial Basis Function) Neural Net-work, Support Vector Machine and Artificial Immune System. This new concept is proposed for the users who access SaaS application through mobile devices.

Biometric cues-based authentication and behaviour-based authentication methods in [27] include face and iris, finger-print, voice/speech etc. [28] make use of gait recognition to detect whether the device is being used by the rightful owner. But the accuracy of result can be affected by footwear, ground surface, carrying load and injuries which post notable impact on a person's walking style. [29] enhance password patterns robustness with an additional security layer which is called touch screen pattern. Authenticating mobile devices users through keystroke analysis is investigated in [30], [31]. But in order to provide reliable scores, keystroke based authentication usually requires a rather long training phase. If the device uses a soft input method instead of hardware keyboard, it will further complicated the issue.

The implicit authentication approach is proposed in [32] which involves learning and behaviour pattern recognition through collecting and analyzing the daily behaviour (phone call, SMS history, browser history, network information etc.) of the user. However, this authentication method requires training phase conforming with the rule the longer, the better which will leave users devices unprotected for a longer period of time. To fully exploit the benefit of implicit authentication, authors in [33] give a brief introduction about how to integrate implicit authentication into TrustCube [34] framework to authenticate the users. TrustCube is an end-to-end infrastructure that offers measurements of requested elements of targeted person, include identity, platform and environment. Thus, the service provider can make informed decision based on the measurement report. The limitation of this framework is that in some circumstance an application may call another service from a different application. The authentication process between two software applications is out of range of current TrustCube framework.

Apart from proper authentication scheme, malware security issues should not be overlooked when it comes to application security. [35] presents a detailed analysis of Android malware. It systematically classifies malware samples from various aspects including installation methods, activation mechanisms and carried malicious payloads. Several solutions have been proposed to tackle malware, for example, static analysis [36], [37] and dynamic behaviour analysis [38] based detection. The advantages of these approaches are that: (1) they tackle detection of unknown malware (i.e. zero-day malicious code) which traditional signature based antivirus scanners fail to detect effectively. (2) For those SaaS applications which are usually accessed by users through mobile devices. The computation can be off-loaded to the SaaS providers sites. However, these approaches are generic to a broad range of mobile malware and not sufficiently focused on SaaS based applications.

To make the best of various anti-virus solutions, cloud-based Anti-virus framework is presented in [39], [40] to provide Anti-virus protection as a service. But it is limited to signature based antivirus products which may not detect new malware quickly enough. The method of server-side malware detection [41] is complementary method which might be used to enhance CloudAV framework. To further enhance the security from the user side, cloud-based smartphone intrusion detection and response system is proposed in [42], [43] by establishing the exact replica of the device in the cloud and a proxy server for recording outbound and inbound traffic. Their advantages are their ability to detect and handle zero-day attack. Recently, IBMs proposed a method to provide a secure virtual machine launcher [44] which helps to prevent mobile devices from accessing software code that has been maliciously or inadvertently modified once encrypted. It is a response to growing demand for securing BYOD in the enterprise against attacks to SaaS applications accessed via mobile devices. It highlights the significance of malware as a threat to SaaS security.

The typical application loopholes and the possible weak-ness brought by user interface and APIs are discussed in section III. To address the web services-related vulnerabilities, [45] proposed a procedure which point out the validation steps required to verify incoming SOAP request, the two validation steps are based on validating incoming messages for conformance to the claimed schema and security policy. Also, [46] proposed a solution which is based on the usage of message structure information (SOAP account) to preserve the integrity of a SOAP account and detect XML rewriting attacks. The advantage of this method is that it deals with the issue of forging SOAP account. However, this method needs to be put in the context of a real world for more detailed performance analysis. [47] presents some solutions according to each type of XML-based attacks. It also recommends that XML security standard should be utilized which are XML signature, XML encryption, XML key management specification, extensible access control markup language and security assertion markup language to extend standard mechanism to tackle the web service security issues. The limitation of these standards is that they cannot deal with XML DoS/DDoS attacks which could take down the entire web server. To address XML-related DoS/DDoS attacks, a solution is proposed in [48] in which a service-oriented traceback techniques along with a back propagation neutral network is used to find the source of attack and further detect and filter these attacks. However this approach is only limited to handle HTTP and XML based DoS and DDoS attacks.

Virtualization-related security issues have been discussed a lot in [8], [13]. However, a concrete scheme detailing how to implement safe virtualization environment in SaaS is not addressed. Nevertheless, they can still be useful to identify vulnerabilities of virtualization in SaaS. Cyberguard [49] is a virtualization security assurance architecture that provides three different kinds of services: virtual machine service, virtual network service and policy-based trust management service. It provides a VMM-based NetApp trusted loading approach along with a multi-level security isolation approach which is based on virtual machine technologies to enforce VM/VMM security. Also, an adaptive security system deployment mechanism for virtual network environment is designed to guarantee virtual network security. The advantages of this method are: (1) capable of preventing from the occurrence of VM escape by intercepting system calls and process behaviour, and (2) provide virtual machine and virtual network isolation. A novel advanced architecture ACPS (Advanced Cloud Protection System) is proposed in [50]. The aim of ACPS is to increase security to cloud resource, but monitoring kernel and middleware integrity method can be used to protect VM. ACPS can be referenced and applied to enforce VM security for SaaS deployment Security. However, this approach does not address virtual network-related security issues and the compromise of hypervisor need to be addressed more in this article. To secure virtualization, various methods [8] have been proposed for platform hardening in virtual systems such as: secure hardware, secure host operating system, secure hypervisor, secure management interfaces, secure virtual machine. Virtual network security issues [9] such as packet sniffing and spoofing threats can be addressed by implementing a novel virtual network model which deploys a virtual firewall between user groups and routing layer. This approach is susceptible to threats originating from malicious insiders if this insider is in the same shared network with the targeted object.

V. DISCUSSION

The analysis of existing solutions in previous section and table I shows that SaaS security encapsulates a wide range of areas and various techniques. The solution which is applicable in one specific area might not be suitable when it moves into SaaS environment. Modifications need to be implemented to adapt the traditional solution to SaaS.

To design a security system for data security in SaaS platform, a variety of security mechanisms should be used to keep users data safe in storage or transmission. Database can classify the data into different groups based on different levels of required confidentiality by users. Different groups here denote that data storage in database can be in strong encrypted form, light encrypted form or in plaintext. This to a large extent minimizes the loss when a hacker successfully hacks into the database. Erasure code could be utilized to transform the file into a longer file so that the original file can be retrieved from a subset of the longer file version. This approach proves to be particularly efficient in distributed database storage system and can relatively save a presentable storage capacity, the latter makes up for the extra storage used when data need to be encrypted in database. Also, homomorphic encryption is an effective means to ensure the confidentiality of data when passed on to a third party for processing or other type of transactions. When a request is sent to database to ask for permission for data access, each code should be verified at the least cost to protect data storage from attacks like SQL-injection.

For data access control issues, SaaS providers must guarantee that information in the database is only accessible to an authorized party. Logging of all the users and inner administrator access to the resource should be recorded in the log management system. At the same time, frequent check of logging system should be implemented and security controls should be in place to prevent log tampering.

For data backup and recovery, at a minimum SAAS providers should be able to provide RAID storage system in case of a disaster, or an agentless backup solution could be adopted. Also, encryption methods such as AES, RC6 should be employed on backup data to prevent the accidental leakage of sensitive information.

Data integrity can be achieved by constraints and transactions in database management system. For the constraints, primary and foreign key constrains are leveraged to insure entity integrity, referential integrity and domain integrity of data at rest in database. Database transactions comply with the ACID (Atomicity, Consistency, Isolation and Durability) rules to ensure data integrity. Meanwhile, other techniques such as checksums, hash function and error-correcting codes or algorithms (Damm algorithm and Luhn algorithm) should be adopted intertwined to achieve a multi-tier data integrity guarantee.

To solve the data security issues during the transfer, strong encryption and mutual authentication mechanism can be applied such as public key infrastructure. In addition, Transport Layer Security (TLS), Secure Socket Layer (SSL) protocol or Internet Protocol Security (IPsec) can be adopted to provide communication security in SaaS according to different Service Level Agreement. To address SaaS security issues from the viewpoint of application security, a high quality SaaS application programming is indispensable and it can help to avoid several security issues. A few of the common best practices for designing SaaS applications are: (1) Design the application to run stateless, with any necessary user and session data stored either on the client side, or in a distributed storage system which can be accessed by any application instance. Statelessness means that each transaction could be handled by one instance as well as any other; A user might perform the transaction concurrently with dozens of instances during a single session, without ever knowing it; (2) Consider asynchronous I/O operations when designing an application; this way the application can perform additional work while waiting for input and output to complete; (3)

Resources pooling such as threads, network connections, and database connections should be aggregated to maximize computing resources and enable application providers to predict resource utilization rates. Apart from the careful design of SaaS applications, proper anti-DoS/DDoS attack mechanisms should also be in place while XML security standards should be jointly utilized to protect the system from XML-based attacks. Additionally, instead of only adopting traditional authentication method such as password and tokens to authenticate between user and application, more advanced and intelligent authentication such as biometric feature and behaviour-based authentication should be merged to enhance the security. In addition, implicit authentication which utilizes the existing human intelligence approach such as RBF (Radial Basis Function) Neural Net-work, Support Vector Machine, proves to be a promising authentication method which still has vast prospect to be exploited. For the future convenience and development, a robust authentication framework should be established which is suitable for the deployment of implicit authentication to securely collect and transfer required information. A server-side malware-detection or intrusion detection system can be deployed also to ensure the security of the device at user side. This can confirm that the application is running on a secure platform.

From the point of view of SaaS providers, to maintaining a high level security on Virtual Machine and Virtual Network is essential. Proper configuration should be set up such as the limit of physical resource every single virtual machine can have; disable unnecessary network and local administrative interfaces to reduce attack surface; check integrity frequently on associated virtual machines. Meanwhile, monitoring system and Intrusion detection system should be deployed to monitor the activities of each virtual machine in order to detect the abnormal behaviours such as VM escape or any type of com-promise of hypervisor. A virtual firewall should be created among virtual machines to provide the usual packet filtering function. A set of security rules or policies can be defined according to different contexts.

TABLE I: SaaS Challenges and Associated Research Work and Solutions

| SaaS Challenges | Possible Solutions |
|---|---|
| **Data Security** | |
| Data Storage | 1. Combine existing data security approaches [16]<br>2. Data storage scheme which utilized the homomorphic token with distributed verification of erasure-coded data [17] |
| Data Access Control | 1. Implement multi-user access policies [18]<br>2. Enfore security policies for data management [19] |
| Data Backup&recovery | 1. Setup agentless backup and recovery [20]<br>2. Backup strategy which data is encoded into segments using error-tolerant encoding scheme [51] |
| Data Integrity | 1. Use protocol like remote file integrity checking [21]<br>2. Implement cryptographic techniques [52] |
| Data Transfer Security | 1. Trusted Third Party [22]<br>2. Secure Sockets Layer [16] |
| **Application Security** | |
| Application Malware | 1. Deploy static malware detection techniques [36], [37]<br>2. Implement behavior-based dynamic malware detection [38], [53]<br>3. Use signature-based anti-virus techniques [39], [40] |
| Application DoS/DDoS | 1. Implement security techniques against XML/HTTP hacking such as<br>   Service-oriented traceback techniques [48]<br>   Artificial intelligence neural network [48], [54] |
| Web Services-related Security Issues | 1. Verifying incoming SOAP request [45]<br>2. Protect SOAP messages against attacks such as XML rewriting [46]<br>3. Applying various XML security standard to handle web service specific nuances and problems [47] |
| Application Access Control | 1. Behavior-based and biometric character-based authentication<br>   Gait based authentication [28]<br>   Implicit authentication based on touch screen pattern [29]<br>   Implicit authentication based on user behavior [32], [33]<br>   Location based authentication [55], [56]<br>   Keystroke based authentication [30], [31]<br>2. Mutual authentication with identity management and session key establishment [25] |
| **Deployment Security** | |
| VM/VMM Security | 1. VMM-based NetApp trusted loading approach along with a multi-level VM security isolation approach [49]<br>2. Malware detection techniques in Virtualised Environments [57]<br>3. Protect VM by monitoring kernel and middleware integrity [50]<br>4. Implement intrusion detection techniques [58]<br>5. Various platform hardening methods [8] |
| Virtual Network Security | 1. Leverage characteristics of 'route' and 'bridge' virtual networking mode [9]<br>2. Various platform hardening methods [8], [49] |

## VI. CONCLUSION

Software as service is facilitating changes to almost every aspect of our modern life. In this article, Software as a service is examined in the context of four different environments: cloud, mobile, SDN and IOT. SaaS promises scalability, lower cost of integration, reduced time to market, easy upgrades and ease of use to perform proof of concepts. However, in order to achieve these benefits, a number of existing challenges must be resolved. We present a discussion of several of these challenges and analyze existing solutions proposed to tackle these challenges. Finally, we propose possible solutions to effectively to deal with the highlighted SaaS security issues.

Our future research vision will focus on two directions to provide confidentiality, integrity and secure data management for SaaS. First, extending authentication techniques. Second, integrating authentication solution with secure resource management in the virtual environment. Finally, a prototype will be implemented to demonstrate the system's feasibility and performance.